\documentclass[%
superscriptaddress,
 amsmath,amssymb,
 aps,
]{revtex4-2}
\usepackage{geometry}
\usepackage{subcaption}
\usepackage{graphicx}
\usepackage{dcolumn}
\usepackage{color}
\usepackage{comment}

 \global\long\def\halfsize{0.45\columnwidth}
\usepackage{bm}
\usepackage{comment} 
\usepackage{hyperref}
\hypersetup{colorlinks=true, citecolor=blue, urlcolor=blue, linkcolor=blue}
\begin{document}
\title{Initial conditions for   Starobinsky inflation in general quadratic gravity}
\author{Daniel M\"uller} 
\email{dmuller@unb.br}
\affiliation{Instituto de F\'{i}sica, Universidade de Bras\'{i}lia, Caixa Postal
04455, 70919-970 Bras\'{i}lia, Brazil}
\author{Alexey Toporensky} 
\email{atopor@rambler.ru}
\affiliation{Sternberg Astronomical Institute, Moscow University, Moscow 119991, Russian Federation}
\begin{abstract}
We consider initial conditions leading to Starobinsky inflation in the general quadratic gravity, where the action of the theory contains one more curvature square 
invariant in addition to $R^2$. We have chosen corresponding coefficients in a way so that the inflationary solution keeps to be stable. Our numerical results show that
despite the configuration of initial conditions in the $(H,R)$ plane, leading to Starobinsky inflation can change considerably from the $R^2$ theory, realization of inflation
does not need a fine-tuning of the initial conditions.
\end{abstract}
\maketitle

\section{Introduction}

Starobinsky inflation \cite{starobinsky1980new}, being one of the oldest inflationary models is still one of the most successful models, regarding ever increasing accuracy of
modern observations. In this theory there is no physical scalar field, and the inflationary behavior is achieved due to the presence of curvature square
corrections $ R^2$ to Hilbert-Einstein action of General Relativity.
 Its predictions match with observed value of primordial perturbations spectral tilt, as well as predicted amplitude of primordial
gravitational waves is well within observational limits. The amplitude of power spectrum fixes the only free constant of the theory, the coupling to curvature
square term in the action. To match with observable value this constant should be chosen to be $\beta= 1.305\times10^9m^{-2}_{\mbox{\tiny P}}$ \cite{Ade:2015lrj} where $m_{\mbox{\tiny P}}=1/\sqrt{8\pi G}$ is the reduced Planck mass. 

On the other hand, the $R^2$ gravity itself is a particular case of more general quadratic gravity where other quadratic curvature invariants do contribute
the action. It is well-known that there are three independent invariants -- namely $R^{ijkl}R_{ijkl}$, $R^{ij}R_{ij}$ and $R^2$. However in $3+1$ dimensions
it is enough to add only one more invariant, since the 3-d one can be expressed through 1-st, 2-nd and the Gauss-Bonnet combination, which is a full
derivative in $3+1$ dimensions and, thus, does not contribute to equations of motion.

Quadratic gravity began with Weyl in $1918$ \cite{weyl1918gravitation}. After many years it was studied by Buchdahl \cite{buchdahl1962gravitational}. Starobinsky inflation is the specific case when there's only the Einsten term and the $R^2$ term in the action, as was found out by Ruzmaikina and Ruzmaikin \cite{ruzmaikina1970quadratic}. Many  articles have analyzed quadratic gravity, see, for example, \cite{tomita1978anisotropic,Berkin:1991nb,Cotsakis:1997ck,Miritzis:2003eu,Barrow:2005qv,Barrow:2006xb,0264-9381-23-9-011,Miritzis:2007yn,COTSAKIS2007341,cotsakis2008slice,Carloni:2007br,Barrow:2009gx} and for a historical review see \cite{schmidt2007fourth}.

We have previously investigated initial conditions for Starobinsky inflation in $R+R^2$ theory for spatially flat \cite{Mishra:2019ymr} and with positive spatial curvature \cite{Muller:2023tsh}. As the next step, the question of Starobinsky inflation in the framework of general quadratic gravity arises naturally. What can be shown easily, is that adding
$R^{ij}R_{ij}$ term does not change properties of isotropic cosmological evolution, since such a theory can be reformulated as a theory containing
the sum of Einstein and Weyl square terms, and Weyl tensor is known to vanish for isotropic metric. However, there are no {\it a priory} reasons why
cosmological shear should vanish (or, at least, be very small) before inflation. So that, investigations of possible impacts of other quadratic terms
to the possibility of entering inflationary regimes make sense for initially anisotropic metrics. 

It is already known through the linearized degrees of freedom that general quadratic gravity has a scalar degree of freedom of mass $m_0=1/\sqrt{6\beta}$, a massless spin 2 field and a spin 2 field with mass $m_2=1/\sqrt{-\alpha}$ \cite{Stelle:1977ry}, see also the appendix in \cite{vanDam:1970vg}. While choosing $\beta>0$ and $\alpha<0$ eliminates the tachyons, it does not eliminate the ghost. The ghost comes with a different sign in the energy of the massive spin 2 field in the linear regime. Such that for sufficiently small perturbations the coupling between these fields is negligible and these fields behave as free fields. On the other hand, by increasing the initial shear at some threshold, the coupling becomes relevant and energy can be transferred from the other fields to the spin 2 massive ghost field which have opposite signs. Linearized energy is conserved and amplitudes of the perturbations can increase. Eventually this mechanism brings the system out of the linear regime and to some singularity.

In our study we restrict ourselves by the case
of zero spatial curvature, so we consider a Bianch I background.

It is known that in the theory with both $ R^{ij}R_{ij}$ and $ R^2$ terms in the action  isotropic solutions
can become unstable with respect to any nonzero shear. However, for inflationary regime this happens only when a simple relation between the coefficients before these terms  is not satisfied (see below).
In the opposite case Starobinsky inflation is a stable regime, and a natural question is to specify initial conditions leading to it. In particular, it is
important to understand if this regime appears naturally or requires some sort of fine-tunning of initial conditions. To answer this question we provide
a numerical analysis of corresponding cosmological dynamics for a case when $\alpha$ is as big as $\beta$ but satisfies the inflation stability condition.

\section{The field equations}

In this work we consider the following Lagrangian density 
\[
{\cal L}=\frac{1}{16\pi G}\left[R+\left(\beta-\frac{1}{3}\alpha\right)R^{2}+\alpha R_{ab}R^{ab}\right].
\]
Metric variations result in the following equations of motion 
\begin{align}
G_{ab}+\left(\beta-\frac{1}{3}\alpha\right)H_{ab}^{(1)}+\alpha H_{ab}^{(2)}=0,\label{field_eq}
\end{align}
where 
\begin{align*}
 & G_{ab}=R_{ab}-\frac{1}{2}Rg_{ab}\\
 & H_{ab}^{(1)}=2g_{ab}\square R-2R_{;ab}+2RR_{ab}-\frac{1}{2}g_{ab}R^{2}\\
 & H_{ab}^{(2)}=\square R_{ab}-R_{;ab}+\frac{1}{2}\square Rg_{ab}+2R^{cd}R_{cbda}-\frac{1}{2}g_{ab}R_{cd}R^{cd}.
\end{align*}
 Stable vaccum Minkowski space requires in this case that $\beta>0$
and $\alpha<0$, \cite{vanDam:1970vg}, \cite{Stelle:1977ry} Using tetrads the metric is 
\[
\tilde{g}_{\alpha\beta}e_{a}^{\alpha}e_{b}^{\beta}=g_{ab}=\mbox{diag}[-1,1,1,1].
\]
As is well known, in this case the connection follows from the covariant
conservation of the above metric and from the zero torsion condition
respectively \cite{stephani2003exact}
\begin{align}
 & \nabla_{a}g_{bc}=0\implies\Gamma_{b\,ca}+\Gamma_{c\,ba}=g_{ab|c}\label{cov_diff_g}\\
 & \nabla_{a}e_{b}-\nabla_{b}e_{a}=[e_{a},e_{b}]=D_{ab}^{c}e_{c},\label{comutador}
\end{align}
where $\nabla_{b}e_{a}=\Gamma_{ab}^{c}e_{c}$ is the connection for
the tetrad base, $D_{ab}^{c}$ is the commutator and $|$ is the directional
derivative, $f_{|c}=e_{c}^{\alpha}\partial_{\alpha}f$. The temporal
component $e_{0}$, also denoted as $n^{a}=(1,0,0,0)$ is geodesic
with zero vorticity and the decomposition of it's covariant derivative
in a trace free and trace parts, 
\[
\nabla_{a}n_{b}=\sigma_{ab}+H\delta_{ab},
\]
for the diagonal shear, 
\[
\sigma_{ab}=\mbox{diag}[0,-2\sigma_{+},\sigma_{+}+\sqrt{3}\sigma_{-},\sigma_{+}-\sqrt{3}\sigma_{-}],
\]
together with (\ref{cov_diff_g}) and (\ref{comutador}) results in
the only non null connection coefficients 
\begin{align*}
 & \Gamma_{11}^{0}=\Gamma_{01}^{1}=-2\sigma_{+}+H & \Gamma_{22}^{0}=\Gamma_{02}^{2}=\sigma_{+}+\sqrt{3}\sigma_{-} &  & \Gamma_{33}^{0}=\Gamma_{03}^{3}=\sigma_{+}-\sqrt{3}\sigma_{-}.
\end{align*}
Remind that in this setting, the connection is not symmetric, for
example $\Gamma_{10}^{0}=\Gamma_{20}^{0}=\Gamma_{30}^{0}=0.$ 
The Riemann tensor is given by 
\[
R_{bcd}^{a}=\Gamma_{bd|c}^{a}-\Gamma_{bc|d}^{a}+\Gamma_{fc}^{a}\Gamma_{bd}^{f}-\Gamma_{fd}^{a}\Gamma_{bc}^{f}-D_{cd}^{f}\Gamma_{bf}^{a},
\]
and the Riemann scalar reads 
\begin{align}
    R=6\left( \sigma_-^2+\sigma_+^2+2H^2+\dot{H}\right) .\label{R_scalar}
\end{align}
In this case the Jacobi identity does not imposes additional conditions
and is identically satisfied by this connection, namely $R_{abcd}+R_{acdb}+R_{adbc}=0.$
 
As is well known Starobinsky-Ruzmaikina-Ruzmaikin inflaton \cite{ruzmaikina1970quadratic} is an
asymptotic solution if the shear goes to zero. In this case the field
equations do not depend on $\alpha$. This is the advantage of the action we use,  since the simple sum $\beta R + \alpha R^{ij}R_{ij}$
in the action would lead to additional term proportional to $\alpha$ even in the absence of shear making the comparison with pure $R+R^2$ gravity more difficult. Now  for zero shear the $00$ field equation
gives
\[
3H^{2}+36\beta H\ddot{H}+108\beta\dot{H}H^{2}-18\beta\dot{H}^{2}=0,
\]
which when $H>>\dot{H}>>\ddot{H}$, 
\[
3H^{2}(1+36\beta\dot{H})\approx0\implies H=const.-\frac{t}{36\beta}.
\]
This last expression is the inflationary Ruzmaikina-Ruzmaikin-Starobinsky 
solution. Linearization around this solution reveals that the frequencies
are 
\begin{align*}
 & &\lambda_{1},\,\lambda_{2}=\left(-\frac{3}{2}+\frac{1}{2}\sqrt{1+\frac{96\beta}{\alpha}}\right)H\begin{aligned} & &\lambda_{3},\,\lambda_{4}=\left(-\frac{3}{2}-\frac{1}{2}\sqrt{1+\frac{96\beta}{\alpha}}\right)H\end{aligned}
\begin{aligned} & &\lambda_{5} \mbox{ to }\lambda_{8}=-3H.\end{aligned}
\end{align*}
Stability requires $H>0$ and that the real parts of the eigenvalues are negative.
We see easily that the conditions  $\beta>0$ and $\alpha<0$ mentioned in the Introduction are enough to  ensures stability of the inflationary solution.
In this article we choose $\alpha=-10\beta$, which is big but within the stability domain.  

\section{Numerical results}
In this section we present our results of numerical integration for the cosmological evolution
of Bianchi I space-time in general quadratic gravity. We fix shear so that $\sigma_-=0$, $\sigma_+$ is specified on plots and set a grid of initial $H$ and $R$ given by \eqref{R_scalar}. Since the shear is fixed, by choosing a value for $R$ corresponds in choosing a value for $\dot H$. After that $\ddot H$ is determined
by the constraint equation \eqref{constraint}. For appropriate initial conditions for inflation we suppose on the order of $60$ e-folds. These initial conditions (marked as black points) in plots made in coordinates
$H$ and $R$. 

\begin{figure}[h]
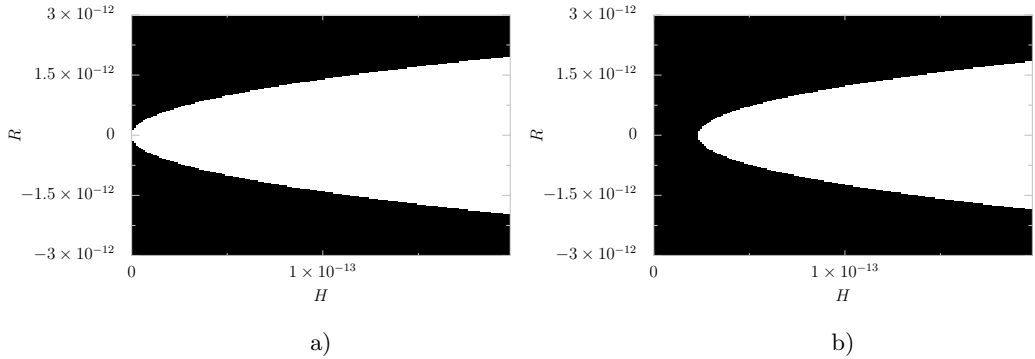

\centering
 \begin{center}
\begin{tabular}{c c} 
       \resizebox{\halfsize}{!}{\input{figure1}} 
      & \resizebox{\halfsize}{!}{\input{figure2}}\\
      a) & b) 
   \end{tabular}
 
    \end{center}
\caption{a) Basin for very small shear  $\sigma_+=1.0\times10^{-10}$. This picture looks almost the same as for exactly zero shear presented in Fig.6a) of \cite{Mishra:2019ymr}.  b) The same for  initially bigger shear, namely  $\sigma_+=1.0\times10^{-8}$. Black points mark initial conditions with sufficient inflation while white points refer to insufficient inflation.}	
\label{small_curv}
\end{figure}
\begin{figure}[h]
\centering
 \begin{center}
\begin{tabular}{c } 
\resizebox{\halfsize}{!}{\input{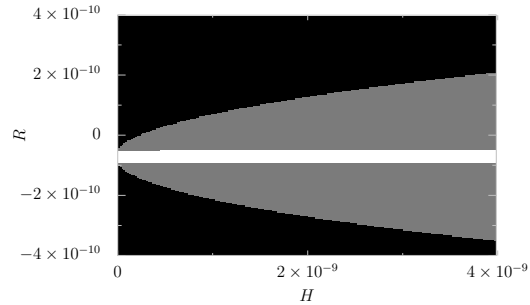}}
   \end{tabular}
 
    \end{center}
\caption{For this plot the shear is increased to   $\sigma_+=2.07\times10^{-6}$. As before, black points mark initial conditions with sufficient inflation, gray points refer to insufficient inflation and white points indicate initial conditions that converge to Big Crunch singularity.}	
\label{begin_singularity}
\end{figure}
\begin{figure}[h]
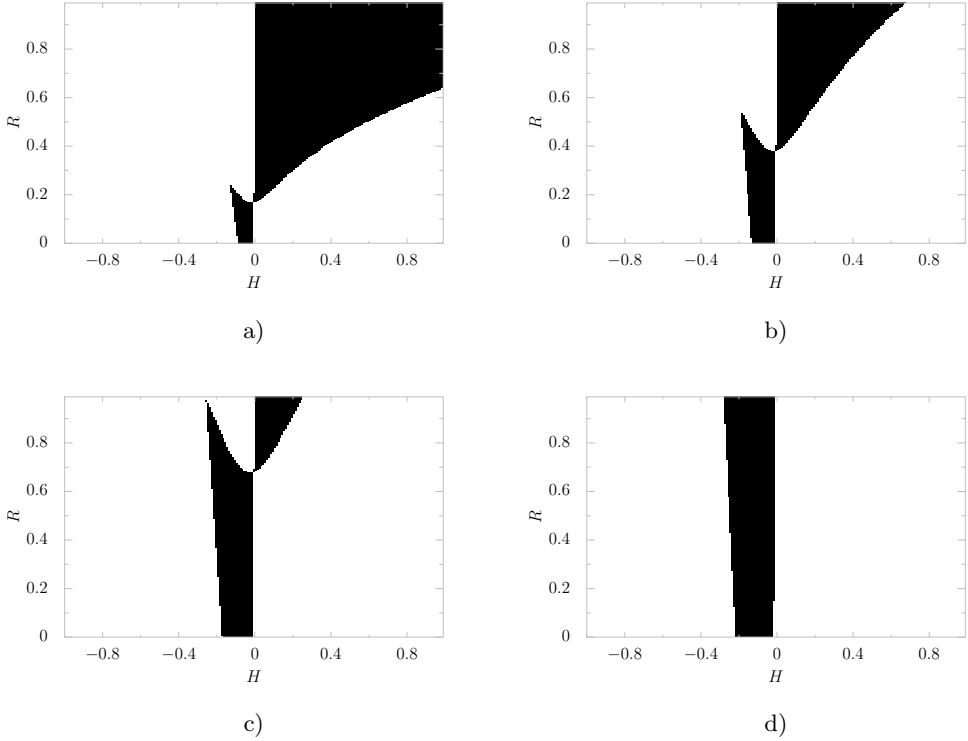

\centering
 \begin{center}
\begin{tabular}{c c} 
\resizebox{\halfsize}{!}{\input{figure4}}
&       \resizebox{\halfsize}{!}{\input{figure5}} \\
      a) & b) \\
      \resizebox{\halfsize}{!}{\input{figure6}}
&       \resizebox{\halfsize}{!}{\input{figure7}} \\
       c) & d)
   \end{tabular}
 
    \end{center}
\caption{We present several basin plots for different initial shear. Each black point marks an initial condition with sufficient inflation, while white points evolve to singularity, Big Crunch. Panel a) $\sigma_+=0.1$; panel b) $\sigma_+=0.15$, panel c) $\sigma_+0.2$ and in panel d) $\sigma_+=0.25$.}	
\label{sub_plnackian}
\end{figure}
We remind the reader that in the absence of shear the condition for Starobinsky inflation to 
occur is roughly $R>0$ with the exception of a thin zone near $R=0$ (where inflation is insufficient) and addition of a small
region with $R<0$ leading to enough inflation. Introduction of non-zero shear changes the situation qualitatively. Very small shear, though, only slightly changes the situation (see Fig.1)
However, for shear about $2 *10^{-6}$ a new feature appears, and some trajectories instead of scalaron oscillations fall into a singularity (possibly, with finer grid they can be remarked
for even smaller shear). 
This means that $H$ changes its sign, becomes negative, and Universe ends in Big Crunch singularity. With increasing shear scalaron oscillations regime disappears completely, moreover,
initial conditions leading to inflation in the absence of shear can lead to a singularity for non-zero shear.

Describing the situation with large shear  we consider only $0<R<1$, $-1<H<1$ region of initial conditions since going beyond $1$ in Planck units is doubtful from obvious physical reasons.
We remark that apart from a turning points in cosmological evolution,
 transitions from contraction to
expansion become possible as well, and some initial conditions with $H<0$ appears to be good for
Starobinsky inflation. 
As it can see from the plots, these two effects work in the opposite directions, always leaving
a room for Starobinsky inflation. With initial shear increases, bigger and bigger parts of initial 
conditions for $H>0$ leading to inflation without shear tend to singularity instead. For shear about
$0.2$ zone of successful inflation moves beyond $H=1$ limit and its physical significance can not be trusted. However, instead a zone of good initial conditions for initially contracting Universe $H<0$ appears and grows with growing shear. For big shear all good initial conditions are those for a contracting Universe which reach Starobinsky inflation going through a bounce.

In any cases, realization of Starobinsky inflation does not need a fine-tuning of initial conditions,
though an actual shape of corresponding zone in initial condition space depends on the value of shear sharply.

\section{Conclusions}
We have considered initial conditions leading to Starobinsky inflation in general quadratic gravity. If the coefficient $\alpha$ before $R^{ij}R_{ij}$ term in the action
belongs to the stability domain, the Starobinsky inflation remains stable with respect to growth of shear variables, however, zone of initial conditions
leading to it can change considerably. We have fixed $\alpha = -10 \beta$, so that $\alpha$ is big, but is located within the stability domain.

As it was anticipated (see Introduction) the influence of $\alpha R^{ij}R_{ij}$ increases with increasing shear. Qualitative difference from the
zero shear  case manifests itself in appearance of turning points of cosmological evolution, the points where $\dot H=0$. For our set-up 
we first remarked the presence of turning points at very small initial shear
of the order of $10^{-6}$ where Big Crunch singularities appear as a finite outcome of the cosmological evolution. However, such a small shear affects only a small
portion of the initial condition space near $(0,0)$ point in coordinates $(H, R)$ and this change is indistinguishable in general plots.

With bigger shear turning points change the picture in two ways. First, the recollaps replaces some parts of initial conditions leading to inflation in the absence
of shear. However, points of bounce appears as well, and some initial conditions with $H<0$ now lead to inflation. This makes the shape of inflation bassin
more complicated for intermediate shear.

Finally, zone of initial condition leading to inflation with initial $H>0$ goes beyond $H<1$, $R<1$ rectangular for big shear. However, there is always a band of
good initial conditions in the initial $H<0$ region. Trajectories from this bassin experience a bounce and reach Starobinsky inflation after it.

It is reasonable to expect that actual form of inflation bassin would depend significantly on the fixed value of $\alpha$ and details of initial shear.  
However, a qualitative result of our study indicates  that these is no need to fix $|\alpha|$ to be much smaller than $\beta$ in the quadratic gravity
(provided  $\alpha$ is within the stability domain) in order to have Starobinsky inflation as a natural outcome of cosmological evolution without need
to fine-tune initial conditions.
\appendix*
\section{Dynamical system}
As already mentioned, this work is for diagonal Bianchi I space. The $00$ field equation \eqref{field_eq} is a constraint that must be initially satisfied 
\begin{eqnarray}
&&E=    \left( 12\dot{\sigma}_+\sigma_+H+12\dot{\sigma}_-\sigma_-H-3{\dot{\sigma}_+}^{2}+6\sigma_+\ddot{\sigma}_+-36{\sigma_+}^{4}+9{\sigma_+}^{2}{H}^{
2}+9{\sigma_-}^{2}{H}^{2}-3{\dot{\sigma}_-}^{2}+6\dot{H}{\sigma_+}^
{2}-36{\sigma_-}^{4}\right.\nonumber\\
&&\left.-72{\sigma_-}^{2}{\sigma_+}^{2}+6\sigma_-\ddot{\sigma}_-
+6{\sigma_-}^{2}\dot{H} \right) \alpha+ \left( -18{\dot{H}}^{2}+72
\dot{\sigma}_+\sigma_+H+108\dot{H}{H}^{2}+72\dot{\sigma}_-\sigma_-H +36H\ddot{H}-54{\sigma_-}^{4}\right.\nonumber\\
&&\left.-108{\sigma_-}^{2}{H}
^{2}-54{\sigma_+}^{4}-108{\sigma_-}^{2}{\sigma_+}^{2}-72\dot{H}{\sigma_+}^{2}
-72{\sigma_-}^{2}\dot{H}-108{\sigma_+}^{2}{H}^{2}
 \right) \beta+3{H}^{2}-3{\sigma_+}^{2}-3{\sigma_-}^{2}.
 \label{constraint}
\end{eqnarray}
On the other hand, the spatial field equations \eqref{field_eq} $11$, $22$ and $33$ can be combined to obtain  
\begin{eqnarray}
&&\dddot{H}=\frac{1}{12\beta}\left[-48\beta \dot{\sigma}_+\sigma_+H+4\alpha \dot{\sigma}_+\sigma_+H-48\beta \dot{\sigma}_-\sigma_-H+4\alpha \dot{\sigma}_-\sigma_-H-3{H}^{2}-3{\sigma_+}^{2}-3{\sigma_-}^{2}-2\dot{H}-54\beta{\dot{H}}^{2} \right.\nonumber\\
&&-\alpha{\dot{\sigma}_+}^{2}-\alpha{\dot{\sigma}_-}^{2}-
18\beta{\sigma_+}^{4}-12\alpha{\sigma_+}^{4}-18\beta{\sigma_-
}^{4}-12\alpha{\sigma_-}^{4}-108\beta \dot{H}{H}^{2}-72
\beta H\ddot{H}+3\alpha{\sigma_+}^{2}{H}^{2}-36\beta
{\sigma_+}^{2}{H}^{2}\nonumber\\
&&+2\alpha \sigma_+\ddot{\sigma}_++2\alpha \dot{H}{\sigma_+}^{2}-24\beta \dot{H}{\sigma_+}^{2}+3\alpha{\sigma_-}^
{2}{H}^{2}-36\beta{\sigma_-}^{2}{H}^{2}-36\beta{\sigma_- }^{2}{\sigma_+}^{2}-24\alpha{\sigma_-}^{2}{\sigma_+}^{2}+2
\alpha \sigma_-\ddot{\sigma}_-\nonumber\\
&&\left.-24\beta{\sigma_-}^{2}\dot{H}+2\alpha{\sigma_- }^{2}\dot{H}-24\beta{\dot{\sigma}_+}^{2}-24\beta{\dot{\sigma}_-}^{2}-24
\beta \sigma_-\ddot{\sigma}_--24\beta \sigma_+\ddot{\sigma}_+\right]
\end{eqnarray}
\begin{eqnarray}
&&\dddot{\sigma}_+=-\frac{1}{\alpha} \left[+6 \alpha H 
\ddot{\sigma}_+-3 \sigma_+H-12 \beta \sigma_+\ddot{H}+4 
\alpha \dot{\sigma}_+\dot{H}-24 \alpha \dot{\sigma}_+{\sigma_+}^{2}+11 \alpha 
\dot{\sigma}_+{H}^{2}-24 \alpha {\sigma_+}^{3}H+6 \alpha \sigma_+
{H}^{3}\right.\nonumber\\
&&-24 \alpha {\sigma_-}^{2}\sigma_+ 
H-8 \alpha {\sigma_-}^{2}\dot{\sigma}_+-12 \beta \dot{\sigma}_+\dot{H}-36 \beta \dot{\sigma}_+{
\sigma_+}^{2}-24 \beta \dot{\sigma}_+{H}^{2}-36 \beta {\sigma_+}^{3}
H-72 \beta \sigma_+{H}^{3}-12 \beta {\sigma_-}^{2}\dot{\sigma}_+\nonumber\\
&&\left.-\dot{\sigma}_+
-16 \alpha \dot{\sigma}_-\sigma_+\sigma_--84 \beta \dot{H}\sigma_+H
-36 \beta {\sigma_-}^{2}\sigma_+H-24 \beta \dot{\sigma}_-\sigma_+ 
\sigma_-+\alpha \sigma_+\ddot{H}+7 \alpha \dot{H}\sigma_+H\right]
\end{eqnarray}
\begin{eqnarray}
&&\dddot{\sigma}_-=    -\frac{1}{\alpha}\left[-\dot{\sigma}_--3 \sigma_- H+4 \alpha \dot{\sigma}_- \dot{H}-8 
\alpha \dot{\sigma}_-{ \sigma_+}^{2}+11 \alpha \dot{\sigma}_-{ H}^{2}-24 
\alpha {\sigma_-}^{2}\dot{\sigma}_--24 \alpha {\sigma_-}^{3} H+6 
\alpha \sigma_-{ H}^{3}\right.\nonumber\\ 
&&+\alpha \sigma_-
 \ddot{H}-12 \beta \dot{\sigma}_- \dot{H}-12 \beta \dot{\sigma}_-{ \sigma_+}^{2}-24 
\beta \dot{\sigma}_-{ H}^{2}-36 \beta {\sigma_-}^{2}\dot{\sigma}_--36 \beta
 {\sigma_-}^{3} H-72 \beta \sigma_-{ H}^{3}-16 \alpha \dot{\sigma}_+ \sigma_+\sigma_-\nonumber\\
 &&\left.+7 \alpha  \dot{H}\sigma_-
 H-24 \alpha { \sigma_+}^{2}\sigma_- H-24 \beta \dot{\sigma}_+
 \sigma_+\sigma_--84 \beta  \dot{H}\sigma_- H-36 \beta { \sigma_+}^{2}
\sigma_- H+6 \alpha  H\ddot{\sigma}_--12 \beta \sigma_- \ddot{H}\right]
\end{eqnarray}
\bibliographystyle{apsrev4-2}
\bibliography{refsR2.bib}
\end{document}